\documentclass[12pt]{iopart}
\usepackage{bm}
\usepackage{graphicx}
\usepackage{epsfig}
\begin{document}

\title[Lagrangian Fuzzy Dynamics of Physical and Non-Physical Systems]
{Lagrangian Fuzzy Dynamics of Physical and Non-Physical Systems}
\author{U. Sandler}
\address{Jerusalem College of Technology (JCT), Jerusalem 91160, Israel.}
\date{\today}

\begin{abstract}
In this paper, we show how to study the evolution of a system, given \emph{imprecise} knowledge about the state of the system and the dynamics laws.  Our approach is based on \emph{Fuzzy Set Theory}, and it will be shown that the \emph{Fuzzy Dynamics} of a $n$-dimensional system is equivalent to Lagrangian (or Hamiltonian) mechanics in a $n+1$-dimensional space. In some cases, however, the corresponding Lagrangian is more general than the usual one and could depend on the action. In this case, Lagrange's equations gain a non-zero right side proportional to the derivative of the Lagrangian with respect to the action.  Examples of such systems are unstable systems, systems with dissipation and systems which can remember their history. Moreover, in certain situations, the Lagrangian could be a \emph{set-valued} function. The corresponding equations of motion then become \emph{differential inclusions} instead of differential equations.  We will also show that the \emph{principal of least action} is a \emph{consequence} of the causality principle and the local topology of the state space and not an independent axiom of classical mechanics.

We emphasize that our adaptation of Lagrangian mechanics does not use or depend on specific properties of the physical system being modeled. Therefore, this Lagrangian approach may be equally applied to \emph{non-physical} systems. An example of such an application is presented as well.
\end{abstract}
\submitto{\JPA}
\maketitle

\section*{Introduction \label{In}}

In experiments with complex systems (for example, living cells), many of the system's parameters remain hidden or out-of-control. This leads to large deviations in experimental results.  As a result, small differences in the \emph{numerical} values of the experimental data lose their significance. Indeed, the state of such a system is better described by a \emph{domain} of points rather than a \emph{single} point in the state space of the system. Moreover, these domains are ``cloud-like'' and do not have crisp \emph{boundaries}. In order to give mathematical meaning to such domains, L. Zadeh introduced the notion of  \emph{Fuzzy Sets} and proposed \emph{Fuzzy Set Theory} and \emph{Fuzzy Logic} \cite{Zadeh 1965}.

Zadeh's basic idea may be illustrated as follows. Assume that the system parameters are described by the variables $\{x_1,x_2,...,x_n\}$, such that each state of the system is represented by a point $\bm{x}=\{x_1,...,x_n\}$ in the abstract space $\mathcal{X}$. An ordinary domain or subset $\mathcal{D}$ of the space $\mathcal{X}$ can be identified with its \emph{Identity function} $\mbox{Id}_\mathcal{D}(\bm{x})$:
\begin{equation}\label{E1.1}
    \mbox{Id}_\mathcal{D}(\bm{x})=\left\{
    \begin{array}{ll}
      1, &  \mbox{ if } \bm{x}\in \mathcal{D}\\
      0, &  \mbox{ otherwise.}  \\
    \end{array}\right.
\end{equation}
A fuzzy set, on the other hand, is identified with its \emph{membership function} $\mu_D(\bm{x})$, where $0\leq\mu_D(\bm{x})\leq 1$. We think of $\mu_D(\bm{x})$ as the \emph{possibility} that the given point $\bm{x}$ belongs to the set $\mathcal{D}$. For $\mu_D(\bm{x})$, we should take a continuous function which tends to zero outside of some region. Then the points far from the region will have almost no possibility of belonging to the set. On the other hand, due to our imprecise knowledge, a point near the border of $\mathcal{D}$ will have intermediate value of possibility of belonging to the domain.

The fuzzy approach enables one to handle the imprecision by operating with pairs $\{x,\mu_{A}(x)\}$ instead of the parameter $x$ by itself. This approach is not new. In stochastic processes and models of quantum particles, for example, we also use probability distributions of the variables' values or a wave function instead of the variables' values by themselves. In our situation however, the imprecision does not have a stochastic or quantum nature. Hence, the membership functions need not possess the properties of a probability distribution.

In what follows, we will need to calculate the membership functions for \emph{composite statements} like ``$X$ is Large \emph{OR} $X$ is Moderate" and ``$X$ is Large \emph{AND} $V$ is Small". In other words, given the membership functions $\mu_{L}(x)$ and $\mu_{M}(x)$ of ``$X$ is Large" and ``$X$ is Moderate", respectively, we need to define membership functions $\mu_{L\vee M}(x)$ and $\mu_{L\wedge M}(x)$. To insure that our fuzzy logical connectives are compatible with common human logic, we will require them to satisfy certain conditions. For example, the possibility that ``X is Large \emph{OR} is Moderate" and the possibility that ``X is Moderate \emph{OR} X is Large" should be equal, reflecting the symmetry of the ``OR" connective. Also, if $X$ is certainly Moderate ($\mu_M(x)=1$), then the possibility that ``X is Large \emph{OR} X is Moderate'' should equal $1$, while if $X$ is certainly not Moderate ($\mu_M(x)=0$), then the possibility that ``X is Large \emph{OR} X is Moderate'' should be equal to the possibility that ``X is Large".

Let us denote the possibility that ``X is W \emph{OR} X is M" by $\mathcal{C}\{\mu_1,\mu_2\}$. Then, in accordance with the above-mentioned properties, $\mathcal{C}\{\mu_1,\mu_2\}$ should satisfy the following:
\begin{eqnarray}\label{E1.2}
    & &  \mathcal{C}\{\mu_1,\mu_2\} = \mathcal{C}\{\mu_2,\mu_1\}, \\ \label{E1.3}
    & &  \mathcal{C}\{\mu, 0\} = \mu,;\;\ \mathcal{C}\{\mu, 1\} = 1, \\ \label{E1.4}
    & &  \mbox{If } \mu_2\leq\mu_3, \mbox{ then }
    \mathcal{C}\{\mu_1,\mu_2\}\leq \mathcal{C}\{\mu_1,\mu_3\}, \\ \label{E1.5}
    & &  \mathcal{C}\{\mathcal{C}\{\mu_1,\mu_2\},\mu_3\} =
    \mathcal{C}\{\mu_1,\mathcal{C}\{\mu_2,\mu_3\}\}=\mathcal{C}\{\mu_1,\mu_2,\mu_3\}.
\end{eqnarray}
Conditions (\ref{E1.4}) and (\ref{E1.5}) reflect the monotonicity and the associativity, respectively, of the ``OR" connective.

Similarly, if we denote the possibility that``$X$ is W \emph{AND} $Y$ is M" by $T[\mu_1;\mu_2]$, we require:
\begin{eqnarray}\label{E1.6}
    & &  T[\mu;0] = T[0;\mu] = 0,\;\ T[1;\mu] =T[\mu;1]= \mu, \\ \label{E1.7}
    & &  \mbox{If } \mu_2\leq\mu_3, \mbox{ then }
    T[\mu_1;\mu_2]\leq T[\mu_1;\mu_3], \\ \label{E1.8}
    & &  T[T[\mu_1;\mu_2];\mu_3] =
    T[\mu_1;T[\mu_2;\mu_3]]=T[\mu_1;\mu_2;\mu_3].
\end{eqnarray}
Mathematical operations satisfying (\ref{E1.2})-(\ref{E1.5}) are well known and were intensively studied during the last decades (see \cite{Butnariu 1994,Mesiar 2005} and references therein). In the mathematics literature, they are called \emph{triangular conorms}, or \emph{t-conorms}, for short. An operation satisfying (\ref{E1.6})-(\ref{E1.8}) (with $T[\mu_1;\mu_2]=T[\mu_2;\mu_1]$) is called a \emph{triangular norm}, or \emph{t-norm}. In Fuzzy Set Theory, t-norms and conorms define the ``intersection" and ``union" of fuzzy sets. This is reasonable because the condition ``$x$ belongs to $S_1$ \emph{AND} $x$ belongs to $S_2$" ($x\in S_1\bigwedge x\in S_2$, for short) defines the intersection $S_1\bigcap S_2$, while ``$x$ belongs to $S_1$ \emph{OR} $x$ belongs to $S_2$" ($x\in S_1\bigvee x\in S_2$) defines the union $S_1\bigcup S_2$. Many examples of t-norms and conorms can be found in \cite{Mizumoto 1989} and \cite{Mesiar 2005}. Practically, the generalized Dubois-Prade t-norm ($T_{DP} $):
\begin{equation}\label{E1.10}
  T_{DP}[\mu_1;\mu_2] = g^{-1}\left(\frac{g(\mu_1)g(\mu_2)}
  {\max(g(\mu_1),g(\mu_2),g(\alpha))}\right),
\end{equation}
where $0<\alpha\leq 1$ and $g(x)$ is a continuous monotonic function with $g(0)=0,g(1)=1$, could be a good choice, because it joins a wide class of \emph{Pseudo-Product t-norms}: $T[\mu_1;\mu_2] = g^{-1}(g(\mu_1)g(\mu_2))$, ($\alpha=1$)  with important \emph{Min t-norm}: $T[\mu_1;\mu_2]=\min[\mu_2;\mu_1]$, ($\alpha=0$).

Note that the \emph{Min} t-norm is the strongest t-norm in the sense that for any t-norm $T[\mu_1;\mu_2]$, we have $T[\mu_1;\mu_2]\leq\min(\mu_1;\mu_2)$. This means that an arbitrary t-norm is as depicted in Fig.~\ref{F1.1}. Of course, there are an infinite number of different t-norms and conorms, but some additional conditions on them could lead to a unique choice of $\mathcal{C}\{\mu_1,\mu_2\}$. For example, let us assume that $\mathcal{C}\{\mu,\mu\}=\mu$. We can write for $0<\mu_2<\mu_1<1$:
\[
    \mu_1=\mathcal{C}\{\mu_1, 0\}\leq \mathcal{C}\{\mu_1,\mu_2\}\leq \mathcal{C}\{\mu_1,\mu_1\}=\mu_1,
\]
and so
\[
    \mathcal{C}\{\mu_1,\mu_2\}=\mu_1 \mbox{ for } \mu_1>\mu_2.
\]
In the opposite case $0<\mu_1<\mu_2<1$, the same arguments lead to:
\[
    \mathcal{C}\{\mu_1,\mu_2\}=\mu_2  \mbox{ for } \mu_2>\mu_1.
\]
Therefore, the only suitable representation of the t-conorm in the case $\mathcal{C}\{\mu,\mu\}=\mu$ is
\begin{equation}\label{E1.11}
    \mathcal{C}\{\mu_1,\mu_2\} = \max\{\mu_1\;\mu_2\}.
\end{equation}
This result will be used in the next section.

Readers who want to find more rigorous mathematics and more information about Fuzzy Sets and Fuzzy Logic can refer to books published during the last decades, in particular to \cite{Mesiar 2005, Introduct 1, Introduct 2, Introduct 3}. The original papers of L.Zadeh \cite{Zadeh 1965}-\cite{Zadeh 2006} are highly recommended as well.

\section{Fuzzy dynamics \label{sFD}}

Consider a system that is moving in some space with coordinates $\bm{x}=\{x_1,...,x_n\}$. We assume that it is not possible to obtain the exact value of the coordinates $x_i$ or of the velocities $V_i$ at any given time $t$. We can say only that there is some possibility that at time $t$, the system is close to the point $\bm{x}$ and its velocity is close to $\bm{V}$. In such a case, the system's movement can be described as follows. If we denote by $\bm{V}',\bm{V}'',\bm{V}''',...$ the possible values of the velocity $\bm{V}$, we can say that:
\begin{itemize}\label{D1}
    \item If, at the time $t+dt$, the system is located in the vicinity of the
    point $\bm{x}$, then at the previous time $t$, the system could be near the point
    $\bm{x}'\approx \bm{x}-\bm{V}'dt$, or near the point
    $\bm{x}''\approx \bm{x}-\bm{V}''dt$, or near the
    point $\bm{x}'''\approx \bm{x}-\bm{V}'''dt$, or ..., and so on, for all possible
    values of the velocity $\bm{V}$.
\end{itemize}
Let us denote the possibility that the system is in a small domain $\Delta_{\bm{x}}$ around the point $\bm{x}$ at the time $t$ by $m(\Delta_{\bm{x}},t)$. We denote the possibility that near the point $\bm{x}$ and at the time $t$ the system's velocity is approximately $\bm{V}$ by $P(\Delta\bm{V}|\Delta_{\bm{x}},t)$. Then the above expression can be symbolically written as
\begin{eqnarray} \nonumber
    m(\Delta_{\bm{x}},t+dt) &=& \mathcal{C}\left\{ T\left[P(\Delta\bm{V}'|\Delta_{\bm{x}'},t);
    m(\Delta_{\bm{x}'},t)\right]; ... \right. \\ \nonumber
    & & \left. ...;T\left[P(\Delta\bm{V}''|\Delta_{\bm{x}''},t);
    m(\Delta_{\bm{x}''},t)\right]; ... \right. \\ \nonumber
    & & \left. ...T\left[P(\Delta\bm{V}'''|\Delta_{\bm{x}'''},t);
    m(\Delta_{\bm{x}'''},t)\right];...\right. \\ \label{E1.13}
    & & ... \left. \mbox{ and so on}\right\}.
\end{eqnarray}
Expression (\ref{E1.13}) is nothing more than the previous natural language expression, written in symbolic form. In order to translate it into an equation of the system's dynamics, we should define mathematical representations of the expressions $m(\Delta_{\bm{x}},t)$, $P(\Delta\bm{V}|\Delta_{\bm{x}},t)$ and the logical connectives $\mathcal{C}\left\{...;...\right\}$ and $T[...;...]$. It is understood that $m(\Delta_{\bm{x}})$ corresponds to some measure of the domain $\Delta_{\bm{x}}$ \footnote{ In accordance with Fuzzy Logic paradigm, $m(\Delta_{\bm{x}})$ can be considered as a \emph{truth value} of the fact that the system is in the domain $\Delta_{\bm{x}}$.}. If we are interested in distances much more than the characteristic size of the domain $\Delta_{\bm{x}}$, it is reasonable to consider a limit where the domain collapses to point:
\begin{equation}\label{E1.14}
    \lim_{\Delta_{\bm{x}}\to\bm{x}}m(\Delta_{\bm{x}},t) = \mu(\bm{x},t).
\end{equation}
Theoretically, there are two cases:
\begin{eqnarray} \label{E1.14a}
  \mu(\bm{x},t) &\neq& 0 \\ \label{E1.14b}
  \mu(\bm{x},t) &\equiv& 0.
\end{eqnarray}
The first one is the main case of our study, while the second one is equivalent to the probabilistic approach to dynamical problems \footnote{In the first case, $\mu({\bm{x}})$ corresponds to the so called \emph{atomic measure} of the domain, while in the second one $m(\Delta_{\bm{x}})$ can be considered as a common additive measure.}. Further, we will assume that $\mu(\bm{x},t)$ and $P(\bm{V};\bm{x},t)$ are continuous, bounded functions: $0\leq \mu,P\leq 1$, where the value $0$ corresponds to the minimal possibility, and the value $1$ to corresponds to the maximal one.

The connectives $\mathcal{C}\left\{\mu_1;\mu_2\right\}$, $T\left[\mu, P\right]$ can be represented by the various t-norms and t-conorms. It is remarkable, however, that the natural properties of the state space's local topology drastically restrict the available choice of the representations of the connective $\mathcal{C}\left\{...;...\right\}$. To demonstrate this, let us consider two nearest-neighbor domains $\Delta_1$, $\Delta_2$ of the system's state space. It is obvious that the possibility that the system is in the joint domain $\Delta_1\bigcup\Delta_2$ is equal to the possibility that it is in the domain $\Delta_1$ \emph{OR} it is in the domain $\Delta_2$. Hence, we can write
\begin{equation}\nonumber
    m\left(\Delta_1\bigcup\Delta_2\right) =
    \mathcal{C}\left\{m(\Delta_1); m(\Delta_2)\right\}.
\end{equation}
Now, if both domains are collapsed to the same point: $\Delta_{1},\Delta_{2} \to \bm{x}$, we have
\[
  m(\Delta_1)\to m(\Delta_2)\to m\left(\Delta_1\bigcup\Delta_2\right) \to \mu(\bm{x}),
\]
which implies that:
\begin{equation}\label{E1.15}
    \mu(\bm{x}) = \mathcal{C}\left\{\mu(\bm{x});\mu(\bm{x})\right\}.
\end{equation}
As shown above, equation (\ref{E1.15}) implies that
\begin{equation}\label{E1.16}
    \mathcal{C}\left\{\mu_1;\mu_2\right\} = \max\{\mu_1\;\mu_2\}.
\end{equation}
This result is crucial for our study, and it is important that this representation for $\mathcal{C}\left\{...;...\right\}$ is dictated by the local topology of the space rather than our mathematical taste, convenience, \emph{etc}. Similar arguments, however, cannot be employed to the connective $T[...;...]$. The reason is that this connective can include membership functions which depend on variables that belong to different spaces. For example, in the expression (\ref{E1.13}), the possibility $m(\Delta_{\bm{x}})$ depends on domain of the system's state space, while the possibility $P(\Delta\bm{V}|\Delta_{\bm{x}},t)$ depends on the domain of its tangential space. In this case, collapsing both of the domains to the same point is impossible, and, therefore, the above-mentioned argumentation becomes invalid. Thus, the explicit form of the t-norm $T[P,\mu]$ remains arbitrary \footnote{In the case (\ref{E1.14b}), however, unique representation of the $AND$ connective can be found by using topological properties of the system's trajectories \cite{MYbook}.}.

Using (\ref{E1.16}), we can rewrite (\ref{E1.13}) as
\begin{equation}\label{E1.17}
    \mu(\bm{x},t+dt) = \sup_{\bm{V}}{T[P(\bm{V};\bm{x},t);\mu(\bm{x}-\bm{V}dt,t)]},
\end{equation}
which is the \emph{Master Equation} of Fuzzy Dynamics \cite{S 1994}-\cite{MYbook}. (Note that (\ref{E1.17}) is a particular case of Zadeh's so-called \emph{Extension Principle} \cite{Zadeh 1975b}). The system's evolution is described by the function $\mu(\bm{x},t)$, which reflects the possibility that the system's variables have the values $x_1,...,x_n$ at the time $t$. The function $\mu(\bm{x},t)$ should be found by solving Eq.~(\ref{E1.17}) with
the initial condition
\begin{equation}\label{E1.18}
    \mu(\bm{x},0) = \mu_0(\bm{x}),
\end{equation}
where $\mu_0(\bm{x})$ is the possibility that the state of the system was $\bm{x}$ at the time $t=0$. The function $P(\bm{V};\bm{x},t)$  is determined by the system's dynamics law \footnote{In the case (\ref{E1.14b}), ``Master-Equation'' of the fuzzy dynamics is equivalent to an ordinary master-equation of the stochastic dynamics \cite{MYbook}
\[
    \rho(\bm{x},t+\epsilon) = \int \mathcal{P}_{\epsilon}(\bm{x},t;\bm{y},t)\rho(\bm{y},t) d^n\bm{y}
\]
}.

\begin{figure}
  \centering
  \includegraphics[height=8cm]{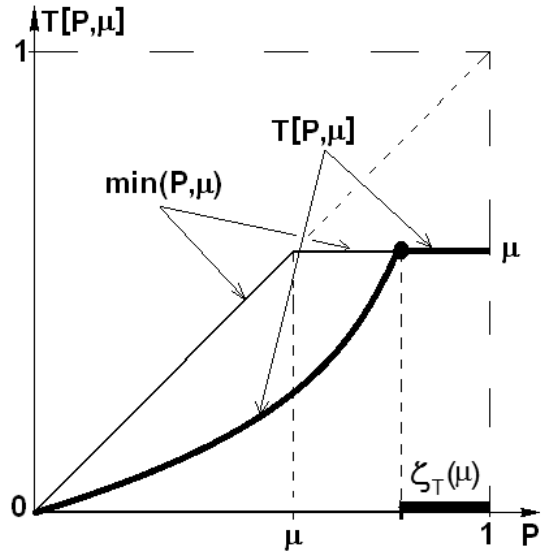}
  \caption{
  Since the function $T[P,\mu]$ is monotonically increasing, $T[P,\mu]\leq\min(P,\mu)$ and $T[1,\mu]=\mu$, $T[P,\mu]$ should be as depicted in the diagram. It is seen that for given $\mu$, the maximum of $T[P,\mu]$ is equal to $\mu$.
  }
  \label{F1.1}
\end{figure}

If we interested in a time interval much more than $dt$, it is reasonable to take the limit $dt\to 0$. To do this, refer to Fig.\ref{F1.1} and note that
\begin{eqnarray}\nonumber
    \sup_{\bm{V}}T[P(\bm{V};\bm{x},t);\mu(\bm{x}-\bm{V}dt,t)] &=& 
    T[P(\bm{V}_m;\bm{x},t);\mu(\bm{x}-\bm{V}_m dt,t)] = \\ \label{E1.19}
    &=& \mu(\bm{x}-\bm{V}_m dt,t),
\end{eqnarray}
where $\bm{V}_m$ is the velocity corresponding to the maximal value of the right side of (\ref{E1.17}). Thus, we can write
\begin{equation}\label{E1.20}
 \mu(\bm{x},t+dt)= \mu(\bm{x}-\bm{V}_m dt,t).
\end{equation}
%000000000000000000
%0000000000000000000
For small $dt$, we can expand $\mu(\bm{x}-\bm{V}_mdt,t)$ with respect to $dt$. In the limit $dt\to 0$, one obtains \cite{S 1994}
\begin{equation}\label{E1.21}
    \frac{\partial\mu}{\partial t} + (\bm{V}_m\cdot\nabla\mu) = 0.
\end{equation}
Note that in order for $\mu(\bm{x},t)-(\bm{V}_m\cdot\nabla\mu)dt$ to be maximal, $(\bm{V}_m\cdot\nabla\mu)$ should be minimal. On the other hand, for $dt\equiv0$, we have
\begin{equation}\label{E1.22}
    T[P(\bm{V}_m;\bm{x},t);\mu(\bm{x},t)]=\mu(\bm{x},t).
\end{equation}
Therefore, $\bm{V}_m(\mu,\nabla\mu;\bm{x},t)$ can be found by minimization of
\begin{equation}\label{E1.23}
    (\bm{V}_m\cdot\nabla\mu)\to \min,
\end{equation}
under the restriction
\begin{equation}\label{E1.24}
    P(\bm{V}_m;\bm{x},t)= \zeta_T(\mu),
\end{equation}
where $\zeta_T$ is a solution of the equation $T[\zeta;\mu]=\mu$. For example, for the Dubois-Prade t-norm, we have
\begin{equation}\label{E1.25}
   \zeta_T(\mu) = \max(\mu,\alpha) + \sin^2(\varphi)[1 - \max(\mu,\alpha)],
\end{equation}
where $\varphi$ is an arbitrary number. (Note that $\zeta_T(\mu)$ doesn't  depend on concrete choice of the function $g(x)$ in (\ref{E1.10})).

Solution of the system (\ref{E1.23}),(\ref{E1.24}) is a well-known problem and can be solved by the method of \emph{Lagrange multipliers}:
\begin{eqnarray}\label{E1.26}
  \lambda\frac{\partial P}{\partial \bm{V}_m} &=& \nabla\mu, \\ \label{E1.26a}
  P(\bm{V}_m;\bm{x},t) &=& \zeta_T(\mu),
\end{eqnarray}
where $\lambda>0$ so that $\bm{V}_m$ will correspond to the minimum of (\ref{E1.23}).

It should be emphasized that the functions  $\mu(\bm{x},t)$ and $P(\bm{V}_m;\bm{x},t)$ cannot be identified with any probability density $\rho(\bm{x},t)$, because they have different mathematical features.  $\mu(\bm{x},t)$  and $P(\bm{V}_m;\bm{x},t)$ are pointwise limited: $\inf(\mu)=\inf(P)=0,\; \sup(\mu)=\sup(P)=1$, while the integral of $\mu(\bm{x},t)$ or  $P(\bm{V}_m;\bm{x},t)$ over all space could be infinite. On the other hand, $\int\rho(\bm{x},t)d^n\bm{x} =1$, while $\rho(\bm{x},t)$ can be infinite at some points \footnote{Actually, $\mu(\bm{x},t)$ is a \emph{function}, while $\rho(\bm{x},t)$ is a \emph{functional}}.

\section{Classical mechanics as fuzzy dynamics in $n+1$-dimensional space \label{CM}}

Consider a dynamical system, whose behavior is described by $n+1$ variables: $n$ coordinates and an additional scalar variable $S$ (\emph{S-variable}). We will assume that our knowledge about the system's location and velocity is imprecise, so that the system's dynamics should be described by its membership function
\begin{equation}\label{E2.0}
    \mu=\mu(S,\bm{x},t),
\end{equation}
and by a function
\begin{equation}\label{E2.0a}
    P=P(L,\bm{V};S,\bm{x},t),
\end{equation}
where $\bm{V}$ is velocity of the system's movement, and $L$ is the rate of change of the $S$-variable. In this case, equations (\ref{E1.21}), (\ref{E1.26})-(\ref{E1.26a}) take the form
\begin{equation}\label{E2.1}
    \frac{\partial\mu}{\partial t} + (\bm{V}\cdot\nabla\mu) + L\frac{\partial\mu}{\partial S}= 0,
\end{equation}
and
\begin{eqnarray}\label{E2.2a}
  \lambda\frac{\partial P}{\partial \bm{V}} &=& \nabla\mu, \\ \label{E2.2b}
  \lambda\frac{\partial P}{\partial L} &=& \frac{\partial \mu}{\partial S}, \\ \label{E2.2c}
  P(\bm{V},L;\bm{x},S,t) &=& \zeta_T(\mu).
\end{eqnarray}
We can solve Eq.(\ref{E2.2c}) with respect to $L$ and obtain
\begin{equation}\label{E2.3}
    L=L(\bm{V},\bm{x},S,\zeta_T(\mu),t).
\end{equation}
Now, substituting (\ref{E2.3}) in (\ref{E2.2c}) and differentiating with respect to $\bm{V}$, one has
\[
   \frac{\partial P}{\partial L}\frac{\partial L}{\partial \bm{V}} +  \frac{\partial P}{\partial\bm{V}}  = 0.
\]
By using (\ref{E2.2a}),(\ref{E2.2b}), we obtain
\begin{equation}\label{E2.5}
    \frac{\partial L}{\partial \bm{V}}=-\frac{\nabla\mu}{\partial_s\mu}.
\end{equation}
Solution of (\ref{E2.5}) with respect to $\bm{V}$ gives
\[
  \bm{V}=\bm{V}\left(-\frac{\nabla\mu}{\partial_s\mu},\bm{x},S,\zeta_T(\mu),t \right).
\]
Finally, substituting (\ref{E2.5}) in (\ref{E2.1}), we obtain
\begin{equation}\label{E2.6}
    \frac{\partial\mu}{\partial t} - H\left(-\frac{\nabla\mu}{\partial_s\mu},\bm{x},S,\zeta_T(\mu),t\right) \frac{\partial\mu}{\partial S}= 0,
\end{equation}
where
\begin{equation}\label{E2.7}
    H=\left(\bm{V}\cdot\frac{\partial L}{\partial \bm{V}} \right) - L.
\end{equation}
Equation (\ref{E2.6}) is a first-order partial differential equation which can be solved by the method of characteristics. The characteristics of equation (\ref{E2.6}) are found from
\begin{equation}\label{E2.8}
  dt = -\frac{d\bm{x}}{\frac{\partial Hw}{\partial \bm{q}}}=-\frac{dS}{\frac{\partial Hw}{\partial w}}=
  \frac{d\bm{q}}{\frac{\partial Hw}{\partial \bm{x}}+\bm{q}\frac{\partial Hw}{\partial \mu}} =
  \frac{dw}{\frac{\partial Hw}{\partial S}+w\frac{\partial Hw}{\partial \mu}}=\frac{d\mu}{\frac{\partial\mu}{\partial t}-Hw},
\end{equation}
where $\bm{q}=\nabla\mu$ and $w=\partial_s\mu$. Equation (\ref{E2.8}) leads to a system of ordinary differential equations:
\begin{eqnarray}\nonumber
  \frac{d\bm{x}}{dt} &=& -w\frac{\partial H}{\partial \bm{q}}, \\ \nonumber
  \frac{dS}{dt} &=& -w\frac{\partial H}{\partial w}-H, \\ \label{E2.9}
  \frac{d\bm{q}}{dt} &=& w\left(\frac{\partial H}{\partial \bm{x}}+\bm{q}\frac{\partial H}{\partial \mu}\right), \\ \nonumber
  \frac{dw}{dt} &=& w\left(\frac{\partial H}{\partial S}+w\frac{\partial H}{\partial \mu}\right), \\ \nonumber
  \frac{d\mu}{dt}&=& 0.
\end{eqnarray}
Introducing a new variable
\begin{equation} \label{E2.9a}
     \bm{p} = -\frac{\bm{q}}{w},
\end{equation}
we can rewrite (\ref{E2.9}) as
\begin{eqnarray} \label{E2.10}
  \frac{d\bm{x}}{dt} &=& \frac{\partial H}{\partial \bm{p}} \\ \label{E2.10a}
  \frac{d\bm{p}}{dt} &=& - \frac{\partial H}{\partial \bm{x}} - \bm{p}\frac{\partial H}{\partial S} \\ \label{E2.10b}
  \frac{dS}{dt} &=& \left(\bm{p}\cdot\frac{\partial H}{\partial \bm{p}}\right) - H = L \\ \label{E2.10c}
  \frac{d\mu}{dt}&=& 0.
\end{eqnarray}
Note that from (\ref{E2.9a}) and (\ref{E2.5}), it follows that
\begin{equation}\label{E2.9b}
    \bm{p} = \frac{\partial L}{\partial \bm{V}}.
\end{equation}
It follows from (\ref{E2.10c}) that $\mu(S(t),\bm{x}(t),t)$ is conserved along the trajectories (\ref{E2.10})-(\ref{E2.10b}). Therefor both $L(\bm{V},\bm{x},S,\zeta_T(\mu_0),t)$ and $H(\bm{p},\bm{x},S,\zeta_T(\mu_0),t)$ depend, in fact, only on the initial value of  $\mu(\bm{x}(0),S(0))=\mu_0$.

Since $\mu(S(t),\bm{x}(t),t)$ is conserved along the trajectories, the system's trajectories are on a surface in the $\{\bm{x},t,S\}$-space, which is defined by equation $\mu(S,\bm{x},t)=\mu_0$ \footnote{In the fuzzy set's literature the set $X_{\alpha} = \left\{ x\in X_{\alpha} : \mu(x)=\alpha \right\}$ is called $\alpha$\emph{-cut} of $\mu(x)$.}. Consider variations $\delta S,\delta\bm{x},\delta t$ on this surface. We have
\[
   \mu(S+\delta S,\bm{x}+\delta\bm{x},t+\delta t) - \mu(\bm{x},S,t)=0,
\]
which leads to
\[
     \frac{\partial\mu}{\partial S}\delta S + (\nabla\mu\cdot\delta\bm{x}) + \frac{\partial\mu}{\partial t}\delta t = 0,
\]
or, using (\ref{E2.6}) and (\ref{E2.9a}), to
\begin{equation}\label{E2.11}
   \delta S =  \left(-\frac{\nabla\mu}{\partial_s\mu}\cdot \delta\bm{x}\right) - \frac{\partial_t\mu}{\partial_s\mu}\delta t
   = (\bm{p}\cdot\delta\bm{x}) - H\delta t.
\end{equation}
This implies that
\begin{eqnarray}\label{E2.11a}
  \frac{\partial S}{\partial\bm{x}} &=& \bm{p} \\ \label{E2.11b}
  \frac{\partial S}{\partial t} &=& -H.
\end{eqnarray}
Note that
\begin{eqnarray}\nonumber
    \frac{\partial H}{\partial p_i}&=&V_i + p_j\frac{\partial V_j}{\partial p_i} - \frac{\partial L}{\partial V_j}
    \frac{\partial V_j}{\partial p_i}=V_i \\  \label{E2.12}
    \frac{\partial H}{\partial x_i}&=& p_j\frac{\partial V_j}{\partial x_i} - \frac{\partial L}{\partial V_j}
    \frac{\partial V_j}{\partial x_i} - \frac{\partial L}{\partial x_i} = - \frac{\partial L}{\partial x_i} \\  \nonumber
    \frac{\partial H}{\partial S}&=&  p_j\frac{\partial V_j}{\partial S} - \frac{\partial L}{\partial V_j}
    \frac{\partial V_j}{\partial S} - \frac{\partial L}{\partial S} = - \frac{\partial L}{\partial S}.
\end{eqnarray}
Thus, using (\ref{E2.12}) and (\ref{E2.9b}), we can rewrite equations (\ref{E2.10a}) and (\ref{E2.10b}) as
\begin{eqnarray} \label{E2.13}
   & & \frac{d}{dt}\frac{\partial L}{\partial\dot{x}_i} - \frac{\partial L}{\partial x_i}=
   \frac{\partial L}{\partial \dot{x}_i}\frac{\partial L}{\partial S} \\ \label{E2.13a}
   & &  S=S_0 + \int_0^t L(\bm{\dot{x}},\bm{x},S,\mu_0,t') dt'.
\end{eqnarray}
Initial conditions for Eqs.(\ref{E2.13})-(\ref{E2.13a}) are
\begin{eqnarray} \label{E2.13i}
  \bm{x}(0) &=&\bm{x}_0, \\ \label{E2.13i1}
  \bm{\dot{x}}(0) &=& \bm{V}\left(-\frac{\nabla\mu_{t=0}}{\partial_s\mu_{t=0}},
  \bm{x}_0,S_0,\zeta_T(\nu_0),0 \right), \\ \label{E2.13i2}
  \mu_{t=0} &=& \mu_0(S_0,\bm{x}_0) = \nu_0,
\end{eqnarray}
where $0\leq \nu_0 \leq 1$ is a fixed number and $S_0(\bm{x}_0)$ should be found from Eq.(\ref{E2.13i2}).

It is easily seen that $S$ is stationary along the trajectories (\ref{E2.13}). Indeed, we have
\[
  \delta S = \int_0^t \left(\frac{\partial L}{\partial\bm{\dot{x}}}\delta\bm{\dot{x}} +
  \frac{\partial L}{\partial\bm{x}}\delta\bm{x} + \frac{\partial L}{\partial S}\delta S\,'\right)dt'.
\]
It follows from (\ref{E2.9b}) and (\ref{E2.11a}) that
\[
  \delta S'= \left(\frac{\partial L}{\partial\bm{\dot{x}}}\cdot\delta\bm{x} \right).
\]
Using (\ref{E2.13}), one obtains
\[
\delta S=(\bm{p}\cdot\delta\bm{x})\left|_0^t\right.,
\]
and so $\delta S=0$ if  $\delta\bm{x}(0)=\delta\bm{x}(t)=0$.
Equations (\ref{E2.11a}),(\ref{E2.11b}) and (\ref{E2.7}) lead to Hamilton-Jacobi equation:
\begin{equation}\label{E2.15i}
    \frac{\partial S}{\partial t} + H\left(\frac{\partial S}{\partial\bm{x}},\bm{x},S,\zeta_T(\mu_0),t\right) = 0
\end{equation}
Note that equations of characteristics of (\ref{E2.15i}) coincide with Eqs.(\ref{E2.10})-(\ref{E2.10b}), so $\mu(\bm{x},S,t)$ is a complete integral of the Hamilton-Jacobi equation.

If $P(L,\bm{V};\bm{x},t)$ doesn't explicitly depend on $S$, then $L(\bm{V},\bm{x},\mu_0,t)$ and $H(\bm{p},\bm{x},\mu_0,t)$
do not depend on $S$, either. In this case, equations (\ref{E2.10})-(\ref{E2.10b}) and (\ref{E2.13}) become the well-known Hamiltonian and Lagrangian equations of classical mechanics, while $S$ in (\ref{E2.13a}) becomes the classical action. As a result, we will call the $S$-variable an ``action," even in the general case. As we can see in (\ref{E2.0}), the action can be considered as added dimension of the state space. In this approach, however, this dimension is not equivalent to the other ones and plays an exclusive role (see, however,  Appendix~\ref{A}).

It should be emphasized that equations (\ref{E2.13}) and (\ref{E2.13a}) were obtained independently by using Eqs.(\ref{E2.1}) and (\ref{E1.26}), which follow from the Master Equation (\ref{E1.17}). This means that stationarity of $S$ (that reflects \emph{principle of least action}) is, in fact, a \emph{consequence} of the causality principle and the local topology of the state space and is \emph{not} an independent axiom of classical mechanics.

\subsection{Uncertainty as an external field \label{CMa}}

Consider a particle which is certainly free far from an origin, but not certainly free in the vicinity of the origin. In this case, the possibility function  $P(L,\bm{V};\bm{x},t)$ could be approximated as
\begin{equation}\label{E2.14}
    P(L,\bm{V};\bm{x}) = \Phi\left( \left[ \frac{L-L_0(V^2)}{ar^{-\alpha}}\right]^2\right),
\end{equation}
where $0\leq\Phi\leq 1$ is any monotonically decreasing function with $\Phi(0)=1$  and $\Phi(\infty)=0$, $L_0(V^2)$ is Lagrangian of the free particle and $r=|\bm{x}|$. The corresponding Lagrangian is
\begin{equation}\label{E2.15}
    L=L_0(V^2) - \frac{\sigma(\mu_0)}{r^{\alpha}},
\end{equation}
where
\[
  \sigma(\mu_0)=\mbox{sign}(\partial_s\mu_0)\sqrt{a\Phi^{-1}(\zeta_T(\mu_0))},
\]
(the sign of $\sigma$ is chosen such that $\lambda$ in (\ref{E1.26}) is positive). We see that uncertainty influences as a ``ghost" field, which disappears for $\zeta_T(\mu_0)=1$ and increases  with decreasing of $\zeta_T$.

\begin{figure}
  \centering
  \includegraphics[height=4.5cm]{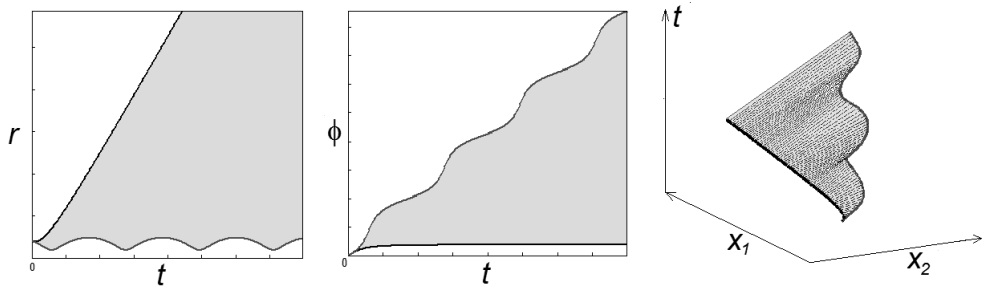} \\
  \caption{Dynamics of the inclusion. Gray area - bundle of the most possible trajectories.}
  \label{Iclus}
\end{figure}

The situation becomes more complicated, however, if we assume that
\begin{equation}\label{E2.16}
    \Phi(x)=\left\{\begin{array}{l}
                          1 \mbox{     if     } 0\leq x \leq l^2, \\
                          \mbox{monotonically decreasing  if    } x>l^2.
                       \end{array}
               \right.
\end{equation}
In this case,
\begin{equation}\label{E2.17}
    L=\left\{\begin{array}{l}
                          L_0(V^2) - alr^{-\alpha} \sin \psi(r)      \mbox{     if    }  \zeta_T(\mu_0)=1, \\
                          L_0(V^2) - \sigma r^{-\alpha}       \mbox{     if     }  \zeta_T(\mu_0)<1,
               \end{array}
       \right.
\end{equation}
where $\psi(r)$ is an arbitrary function of $r$. The Lagrangian (\ref{E2.17}) is a set-valued function because it corresponds to a set of functions and not to a unique function as in (\ref{E2.15}). This means that Eqs.(\ref{E2.13}), with Lagrangian (\ref{E2.17}), and (\ref{E2.10})-(\ref{E2.10b}), with corresponding Hamiltonian (\ref{E2.7}), become \emph{differential inclusions} instead of differential equations (see \cite{Aubin 1990}-\cite{Hullermeir 1997} and references therein for more information about differential inclusions). Solving differential inclusions is more complicated than solving differential equations because inclusions describe the dynamics of a set rather than the dynamics of a point. Fortunately, in our particular case, the solution of inclusion with Lagrangian (\ref{E2.17}) can be found in a simple way by considering (\ref{E2.17}) in polar coordinates
\[
    x_1 = r\cos(\phi), \;\;\; x_2 = r\sin(\phi).
\]
For small velocities, we can write
\begin{equation}\label{E2.18}
    L = \frac{m}{2}(\dot{r}^2 + r^2\dot{\phi}^2) - \frac{al\sin \psi(r)}{r^{\alpha}}.
\end{equation}
Since $\phi$ is a cyclic variable, we have
\begin{equation}\label{E2.19}
    \frac{\partial L}{\partial \dot{\phi}} = mr^2\dot{\phi} = M =const.
\end{equation}
Hence,
\begin{equation}\label{E2.20}
    L = \frac{m\dot{r}^2}{2} + \frac{M^2}{2mr^2} - \frac{al\sin \psi(r)}{r^{\alpha}}.
\end{equation}
The Lagrangian (\ref{E2.20}) does not depend on time. Therefore, the energy
\begin{equation}\label{E2.21}
    E = \left(\bm{V}\cdot\frac{\partial L}{\partial \bm{V}}\right) - L =
    \frac{m\dot{r}^2}{2} + \frac{M^2}{2mr^2} + \frac{al\sin \psi(r)}{r^{\alpha}}
\end{equation}
is conserved. The Lagrangian (\ref{E2.20}) leads to the equations of motion:
\begin{eqnarray}\label{E2.22}
 m \frac{d^2r}{dt^2} &=&  - \frac{M^2}{mr^3} +  \frac{ al[\alpha\sin \psi(r) - r\psi '(r)\cos \psi(r)]}{r^{\alpha+1}}, \\ \label{E2.22a}
  \frac{d\phi}{dt} &=& \frac{M}{mr^2}.
\end{eqnarray}
These equations can be easily solved in an implicit form:
\begin{eqnarray} \label{E2.23}
  t &=& \sqrt{\frac{m}{2}}\int_{r_0}^r \left(E -  \frac{M^2}{2mr^2} - \frac{al\sin \psi(r)}{r^{\alpha}} \right)^{-\frac{1}{2}} dr, \\ \label{E2.23a}
  \phi &=& \frac{M}{\sqrt{2m}}\int \left(E -  \frac{M^2}{2mr^2} - \frac{al\sin \psi(r)}{r^{\alpha}} \right)^{-\frac{1}{2}}\frac{dr}{r^2}  + const.
\end{eqnarray}
Since the solution (\ref{E2.23})-(\ref{E2.23a}) depends on an arbitrary function $\psi(r)$, it describes a set of equally possible trajectories rather than a single trajectory as in the usual (non-fuzzy) case. In order to understand and to interpret the evolution of inclusion, we need to know the border of the set (\ref{E2.23})-(\ref{E2.23a}). In general, this is a nontrivial and complicated task, but in our case, this border can be found quite simply. Indeed, it is seen from (\ref{E2.23}) that its upper and lower borders correspond to $\psi(r)=\mp \pi/2$, respectively. The solution of inclusion (\ref{E2.23})-(\ref{E2.23a}) for $\alpha=1$ is shown in Fig.~\ref{Iclus}.
%00000000000000000000000000000000000
%00000000000000000000000000000000000

\subsection{Cost of memory \label{CMb}}

Consider now the general case, in which the velocity of a system is determined not only by the system's current state, but by its action $S$ as well. This means that the possibility of values of the system's velocity depends on $S$. Hence, in accordance with (\ref{E2.3}), the Lagrangian and the Hamiltonian depend on $S$ as well. We will call such a Lagrangian an \emph{S-Lagrangian}.  It follows from (\ref{E2.13a}) that the $S$-variable depends on a system's history. Therefore, such a system should ``remember" its history. This memory, however, has a certain cost.

Consider a closed system consisting of several particles. In this case, equations (\ref{E2.10})-(\ref{E2.10b}) have the form
\begin{eqnarray} \label{E2.24}
  \frac{d\bm{x}^i}{dt} &=& \frac{\partial H}{\partial \bm{p}_i}, \\ \label{E2.24a}
  \frac{d\bm{p}_i}{dt} &=& - \frac{\partial H}{\partial \bm{x}^i} - \bm{p}_i\frac{\partial H}{\partial S}, \\ \label{E2.24b}
  \frac{dS}{dt} &=& (\bm{p}_i\cdot\bm{V}^i) - H.
\end{eqnarray}
Since the system is closed, we have
\[
   \sum_i \frac{\partial H}{\partial \bm{x}^i} = 0.
\]
Setting $\bm{P}=\sum_i \bm{p}_i$, we obtain
\begin{equation}\label{E2.25}
    \frac{d\bm{P}}{dt} =  - \frac{\partial H}{\partial S}\bm{P},
\end{equation}
and
\begin{equation}\label{E2.26}
    \frac{dH}{dt} = \frac{\partial H}{\partial t} + \frac{\partial H}{\partial \bm{x}^i}\frac{d\bm{x}^i}{dt} +
    \frac{\partial H}{\partial \bm{p}_i}\frac{d\bm{p}_i}{dt} + \frac{\partial H}{\partial S}\frac{dS}{dt} =
    \frac{\partial H}{\partial t} - H\frac{\partial H}{\partial S}.
\end{equation}
This means that even in closed systems with a time-independent Hamiltonian ($\partial H/\partial t=0$), neither total moment nor energy are conserved \footnote{On the other hand for some exotic S-Hamiltonian:
\[
   H=\sum_{k=0}^N h_k(\bm{x},\bm{p},t)S^k
\]
with
\[
   \frac{\partial h_k}{\partial t} = \sum_{q=1}^{k+1} q h_q h_{k+1-q}
\]
the energy is conserved.}
(however, the notions of ``energy", ``moment"  and ``action" of systems with an S-Lagrangian might be different from the usual notions of energy, moment and action (see Appendix \ref{D})). This result is reasonable, because the term $\partial H/\partial S$ in (\ref{E2.24a}) can be considered as an effective friction coefficient. Note that for closed systems with a time-independent S-Hamiltonian, the quantity $\bm{P}/H$ is conserved:
\begin{equation}\label{E2.27}
    \frac{d}{dt}\left(\frac{\bm{P}}{H}\right) = 0.
\end{equation}
It should be noted, that for the systems with an $S$-Lagrangian
\begin{equation}\label{E2.27}
    S(\bm{x}_2,t_2;\bm{x}_1,t_1) \neq S(\bm{x}_2,t_2;\bm{y},t') + S(\bm{y},t';\bm{x}_1,t_1).
\end{equation}

\section{Lagrangian mechanics of non-physical system \label{NonPhysSys} }

Consider an organism which acquires a particular resource $S$ by moving on a surface. It spends this resource in order to maintain the system's activity (in our case, its movement). In addition, the acquired resource spontaneously decays. During the time $dt$, the system obtains an amount of the resource equal to
\begin{equation}\label{E3.1}
    dS = \varrho(\bm{x}) |d\bm{x}| - f_1(|\bm{v}|)dt - f_2(S)dt,
\end{equation}
where $\bm{v}$ is the velocity, $\varrho(\bm{x})$ is proportional to the density of the resource, $d\bm{x}$ is the path of the system during the time $dt$, $f_1(|\bm{v}|)$ is the rate of spending of the resource in maintaining the system's activity, and $f_2(S)$ is rate of spontaneous decay. Equation (\ref{E3.1}) admits different semantics. For example, it could be a simple model of real biological systems such as a caterpillar on a plant or a whale in a plankton field. It could also model the selling of products. In this case, $S$ is income, $\varrho(\bm{x})$ describes the distribution of the buyers, $f_1$ is travel and other expenses, and $f_2$ is the tax obligation.
%00000000000000000000000000000000000
\begin{figure}
  \centering
  \includegraphics[width=14cm]{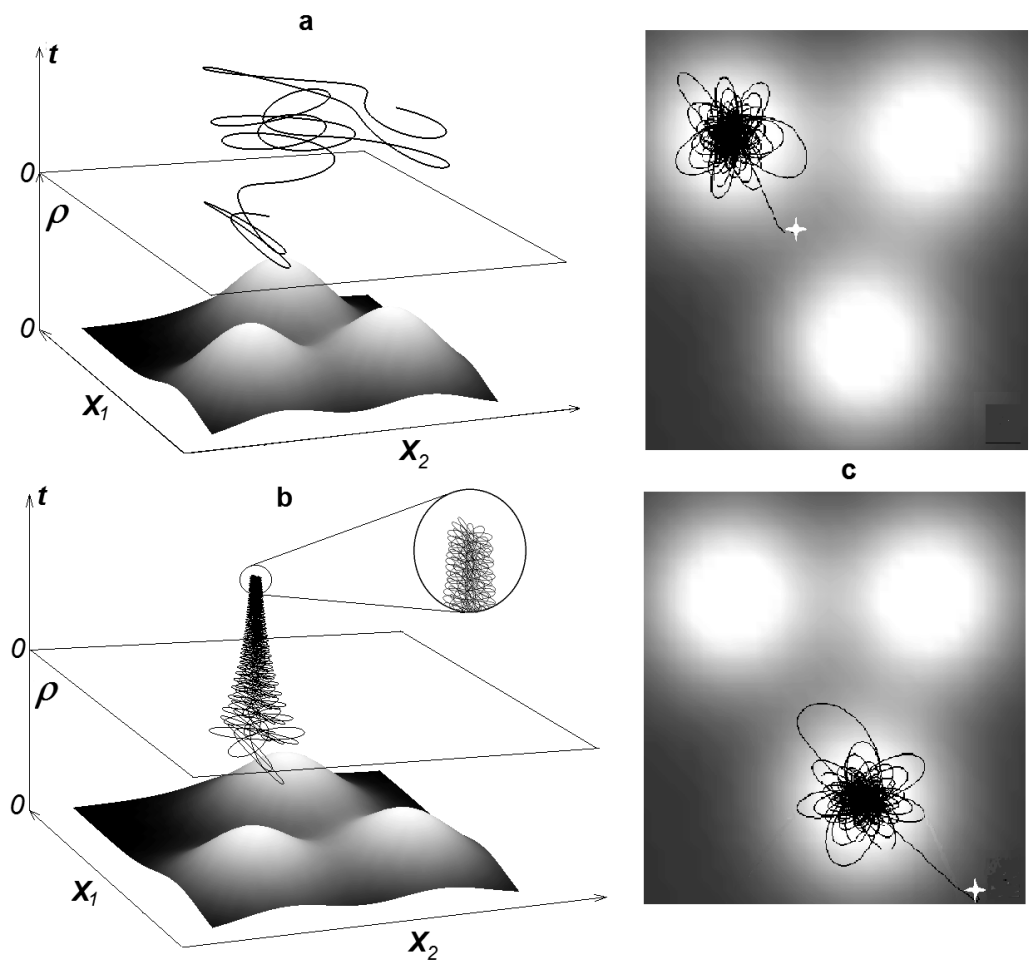} \\
  \caption{Dynamics of the system (\ref{E3.3})-(\ref{E3.3a}): \textbf{a}(upper part) - the system without memory ($\gamma=0$).
  \textbf{b}(upper part) - the system with memory ($\gamma=5$).
  Lower parts of \textbf{a} and \textbf{b} show distribution of the resource.
  \textbf{c} - trajectories for $\gamma=5$ and different initial positions (stars).}
  \label{NonSys1}
\end{figure}
%00000000000000000000000000000000000
%00000000000000000000000000000000000
\begin{figure}
  \centering
  \includegraphics[width=8cm]{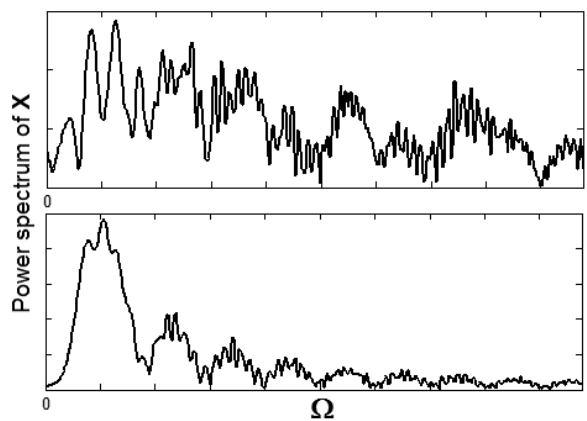} \\
  \caption{Power spectrum of $\bm{x}(t)$, derived from (\ref{E3.3}). \textbf{a}:  $\gamma=0$,
  \textbf{b}:  $\gamma=5$.}
  \label{PSpect}
\end{figure}
%00000000000000000000000000000000000
For small $\bm{v}$ and $S$, we can expand $f_1(|\bm{v}|)$ and $f_2(S)$ with respect to $\bm{v}$ and $S$:
\begin{eqnarray*}
  f_1(|\bm{v}|) &\simeq& \frac{m \bm{v}^2 }{2}, \\
             f_2(S) &\simeq& \gamma S,
\end{eqnarray*}
where we have included the linear term of $f_1(|\bm{v}|)$ in the first term in (\ref{E3.1}). It follows from Eq.(\ref{E3.1}) that in this case, the most possible  $S$-Lagrangian of the system ($P(\bm{v},L_{mp};\bm{x},S)=1$) can be written as
\begin{equation}\label{E3.2}
    L_{mp} = m\left(\rho(\bm{x})v - \frac{v^2}{2}\right) - \gamma S,
\end{equation}
where $v=|\bm{v}|$ and $\rho = \varrho /m$. In accordance with (\ref{E2.13})-(\ref{E2.13a}), the equations of motion for the most possible trajectories ($\mu=1$) are (see Appendix \ref{B}):
%00000000000000000000000000000000000
\begin{figure}
  \centering
  \includegraphics[width=10cm]{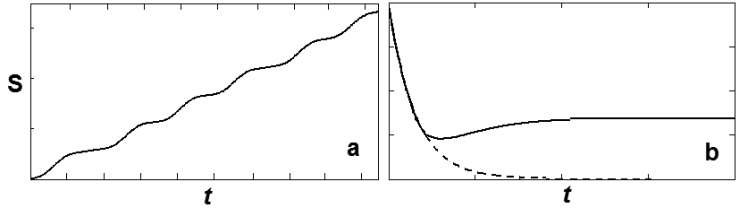} \\
  \caption{Dynamics of the acquired resource. \textbf{a}: The system without memory and without decay of the acquired resource ($\gamma=0$). \textbf{b}: Solid line - the system with memory and with decay of the resource ($\gamma=5$),
  dashed line - the system without memory, but with decay of the resource.}
  \label{Sdyn}
\end{figure}
%00000000000000000000000000000000000
\begin{eqnarray} \label{E3.3}
         \frac{d\bm{x}}{dt}&=& \bm{v}, \\ \label{E3.3a}
         \frac{d\bm{v}}{dt}&=& \frac{v^2}{\rho(\bm{x}) - v}\left(\nabla\rho - \bm{n}(\bm{n}\cdot\nabla\rho)\right) +
         \gamma\frac{\rho(\bm{x}) - v}{v}\bm{n}, \\ \label{E3.3b}
         \frac{dS}{dt}&=& m\left(\rho(\bm{x})v - \frac{v^2}{2}\right) - \gamma S,
\end{eqnarray}
where  $\bm{n}=\bm{v}/v$. We leave the analysis of these equations to Appendix \ref{C} and present here the numerical solutions of (\ref{E3.3})-(\ref{E3.3b}) for different $\gamma$ in a``random" environment such as
\[
    \rho(\bm{x}) = \sum_k A_k \exp{-\left(\frac{\bm{x}-\bm{c}_k}{\sigma_k} \right)^2},
\]
where $A_k,  \bm{c}_k, \sigma_k$ are constants.

It is seen in Fig. \ref{NonSys1}(a) that a system without memory ($\gamma=0$) demonstrates a random walk on the surface (see Fig. \ref{PSpect}(a)). The system with memory ($\gamma>0$), on the other hand, very quickly finds the place to maximize the enriching of its resources and enters this place (see Fig. \ref{NonSys1}(b,c)). The power spectrum of the trajectory (Fig. \ref{PSpect}(b)) implies that in the last case, the system's trajectory becomes a strange attractor, which is concentrated in the resource enriching area.

It should be emphasized that we did not introduce any special equipment for memorization of the system's history. In fact, the dependence of the Lagrangian on the ``action"  (acquired resource in our case) itself creates the system's memory.

Interestingly, the considered model is an example of ``rational behavior without mind." It is seen in Fig. \ref{Sdyn} that in the absence of decay of the resource ($\gamma=0$), the random walk is an effective behavior.  However, if we omit the ``memory term" in the equations of motion (right-hand side of equation (\ref{E2.13})), but keep the decay of the resource in the equation for $S$, (simply set $\gamma=0$ in equation (\ref{E3.3a}), but keep it in equation (\ref{E3.3b})), the random walk will be ineffective, while the strange attractor in the vicinity of the resource-enriching area becomes an optimal strategy.

\section{Quantum mechanics of systems with $S$-Lagrangian \label{QM}}

This Section does not directly relate to fuzzy dynamics, but rather exploits the notion of $S$-Lagrangian introduced above. Consider a quasi-particle with Lagrangian
\begin{equation}\label{E4.1}
    L(\bm{\dot{x}},\bm{x},S) = \frac{m\bm{\dot{x}}^2}{2} - U(\bm{x}) - \gamma S = L_c(\bm{\dot{x}},\bm{x}) - \gamma S,
\end{equation}
where $\gamma >0$ is a constant and $S$ is an action. If, at time $t=0$, the quasi-particle's state was $\psi(\bm{x}_0,0)$, then at time $t$, the state of the particle is
\begin{equation}\label{E4.2}
    \psi(\bm{x},t) = \int d^n\bm{x}_0 G_F(\bm{x},t;\bm{x}_0,0)\psi(\bm{x}_0,0)
\end{equation}
where $G_F(\bm{x},t;\bm{x}_0,0)$ is Feynman's propagator \cite{F}
\[
    G_F(\bm{x},t;\bm{x}_0,0) = \int \mathcal{D}\bm{x}(t) \exp\frac{i}{\hbar}\int_0^t L(\bm{\dot{x}'},\bm{x'},S')dt'
\]
where $\int \mathcal{D}\bm{x}(t)$ is a path integral over all trajectories from $\bm{x}_0$ to $\bm{x}$. It is useful to consider a more general propagator
 \begin{equation}\label{E4.3}
    G_{\alpha}(\bm{x},t;\bm{x}_0,0) = \int \mathcal{D}\bm{x}(t) \exp\frac{i\alpha}{\hbar}\int_0^t L(\bm{\dot{x}'},\bm{x'},S')dt'
\end{equation}
with initial condition:
\[
     \lim_{t\to 0} G_{\alpha}(\bm{x},t;\bm{x}_0,0) = \delta(\bm{x}-\bm{x}_0).
\]
$G_{\alpha}(\bm{x},t;\bm{x}_0,0)$ depends on an additional parameter $\alpha$ and coincides with Feynman's propagator for $\alpha=1$.
Consider a small time interval near the time $t$. It follows from (\ref{E2.10b}) and (\ref{E4.1}) that
\begin{eqnarray}\nonumber
     S(\bm{x},t;\bm{x}_0,0) &=& \int_{t-\varepsilon}^t e^{\gamma(t' - t)} L_c(\bm{\dot{x}}(t'),\bm{x}(t'))dt + e^{-\gamma\varepsilon}S(\bm{x}-\bm{\eta},t-\varepsilon;\bm{x}_0,0)  -  \\ \label{E4.4}
     - S_0\left( 1 - e^{-\gamma \varepsilon}\right) &\simeq& \Delta S + (1-\gamma\varepsilon)S(\bm{x}-\bm{\eta},t-\varepsilon;\bm{x}_0,0)
      - \gamma \varepsilon S_0,
\end{eqnarray}
where $ S_0$ is a constant, $\bm{\eta}$ is the shift of the particle during the time $\varepsilon$ and we have denoted
\[
   S(\bm{x},t;\bm{x}_0,0) = \int_0^t L(\bm{\dot{x}'},\bm{x'},S')dt'.
\]
For small $\varepsilon$ we have
\begin{eqnarray*}
    \Delta S &=& \int_{t-\varepsilon}^t e^{\gamma(t' - t)} L_c(\bm{\dot{x}}(t'),\bm{x}(t'))dt =
    \int_0^\varepsilon e^{-\gamma\theta} L_c(\bm{\dot{x}}(t-\theta),\bm{x}(t-\theta)) d\theta = \\
    &=& \int_0^\varepsilon (1-\gamma\theta)\left[ L_c(t) - \theta\frac{dL_c}{dt} \right]d\theta  + o(\varepsilon^3) \simeq \\ 
    &\simeq&  \left( \varepsilon - \frac{\gamma\varepsilon^2}{2}\right)L_c(\bm{\dot{x}}(t),\bm{x}(t)) - \frac{\varepsilon^2}{2}
    \frac{dL_c}{dt}.
\end{eqnarray*}
Since $\varepsilon$ is small, trajectories during  the time $\varepsilon$ can be considered to be classical \cite{F}, so we can write
\[
    \frac{dL_c}{dt} = \bm{\dot{x}}\left(m\bm{\ddot{x} - \nabla U} \right) = -2(\bm{\dot{x}}\cdot\nabla U) - \gamma m\bm{\dot{x}}^2,
\]
where we used (\ref{E2.13}) with Lagrangian (\ref{E4.1}). Finally, we obtain
\begin{equation}\label{E4.5}
    \Delta S = \frac{m\bm{\eta}^2}{2\,\varepsilon} - \varepsilon U(\bm{x}) + \frac{\gamma m}{4}\bm{\eta}^2 +o(\varepsilon\eta,\varepsilon^2)
\end{equation}
By using (\ref{E4.5}), we can rewrite (\ref{E4.3}) in the form
\begin{eqnarray*}
  G_{\alpha} &\simeq& \frac{1}{A}\int_{-\infty}^\infty d^n \bm{\eta}\, e^{\frac{i\alpha}{\hbar}
  \left( \frac{m\bm{\eta}^2}{2\,\varepsilon} - \varepsilon U(\bm{x}) + \frac{\gamma m}{4}\bm{\eta}^2 \right)}
     \int \mathcal{D}\bm{x}(t)\, e^{\frac{i\alpha(1-\gamma\varepsilon)}{\hbar}S(\bm{x}-\bm{\eta},t-\varepsilon;\bm{x}_0,0)} \simeq \\
     &\simeq& \frac{1}{A}\int_{-\infty}^\infty d^n \bm{\eta} e^{\frac{i\alpha m\bm{\eta}^2}{2\hbar\varepsilon}}
    \left[ 1 - \frac{i\alpha\varepsilon}{\hbar} U(\bm{x}) + \frac{i\alpha\gamma m}{4\hbar}\bm{\eta}^2 \right]
   G_{\alpha-\alpha\gamma\varepsilon}(\bm{x}-\bm{\eta},t-\varepsilon;\bm{x}_0,0).
\end{eqnarray*}
Expanding $G_{\alpha-\alpha\gamma\varepsilon}(\bm{x}-\bm{\eta},t-\varepsilon;\bm{x}_0,0)$ with respect to $\varepsilon$ and $\bm{\eta}$ and integrating over $\bm{\eta}$, we obtain
\begin{equation}\label{E4.6}
    i\hbar\frac{\partial G_\alpha}{\partial t} = - \frac{\hbar^2}{2m\alpha}\nabla^2G_\alpha + \alpha U(\bm{x})G_\alpha
    -  i\hbar\gamma\alpha\frac{\partial G_\alpha}{\partial \alpha} -  i\hbar\frac{n\gamma}{4}G_\alpha.
\end{equation}
The term $\partial G_\alpha/\partial \alpha$ has a clear physical meaning. Indeed, it is follows from (\ref{E4.3}) that
\begin{eqnarray} \nonumber
  i\hbar\frac{\partial G_\alpha}{\partial \alpha}|_{\alpha =1}
  &=& \frac{i\hbar n}{2}G_F  -\int \mathcal{D}\bm{x}(t)\,S(\bm{x},t;\bm{x}_0,0)\, e^{\frac{i}{\hbar}S} = \\ \nonumber
  &=& \frac{i\hbar n}{2}G_F  - \frac{\int \mathcal{D}\bm{x}(t)\,S(\bm{x},t;\bm{x}_0,0)\,
  e^{\frac{i}{\hbar}S} }{\int \mathcal{D}\bm{x}(t)\, e^{\frac{i}{\hbar}S}}G_F = \\ \label{E4.8a}
  &=& \frac{i\hbar n}{2}G_F -  \mathcal{S}_F  G_F
\end{eqnarray}
where we take into account that $\mathcal{D}\bm{x}(t)$ in (\ref{E4.3}) depends on $\alpha$ as well. Thus, for $\alpha =1$, Eq.(\ref{E4.6}) takes the form
\begin{equation}\label{E4.6a}
    i\hbar\frac{\partial G_F}{\partial t} = - \frac{\hbar^2}{2m}\nabla^2G_F +  U(\bm{x})G_F +
    \gamma\mathcal{S}_F G_F -  i\hbar\frac{3n\gamma}{4}G_F = \hat{H}G_F -i\hbar\frac{3n\gamma}{4}G_F ,
\end{equation}
where $\hat{H}$ is the Hamiltonian corresponding to the Lagrangian (\ref{E4.1}):
\[
   \hat{H} = - \frac{\hbar^2}{2m}\nabla^2 +  U(\bm{x}) + \gamma \mathcal{S}_F(\bm{x},t)  = \hat{H}_c +\gamma \mathcal{S}_F.
\]
Solution of (\ref{E4.6}) can be written in the form
\begin{equation}\label{E4.7}
    G_\alpha = e^{-\frac{3n\gamma t}{4}}g_\alpha(\bm{x},t;\bm{x}_0,0),
\end{equation}
where
\begin{equation}\label{E4.8}
    i\hbar\frac{\partial g_\alpha}{\partial t}  = \hat{H} g_\alpha.
\end{equation}
Hence, the Lagrangian (\ref{E4.1}) describes an unstable quasi-particle with life-time $\tau \sim \gamma^{-1}$.

For free quasi-particle ($U(\bm{x})=0$) equation (\ref{E4.6}) can be easily solved (see Appendix~\ref{Q})
\begin{equation}\label{E4.9}
    G_{\alpha}(\bm{x},t;\bm{x}_0,0) = \left( \frac{\gamma m\alpha}{4\pi i\hbar\sinh(\gamma t/2)}\right)^{n/2}
    \exp\frac{i\alpha}{\hbar}\left(\frac{\gamma e^{-\gamma t}m(\bm{x} - \bm{x}_0)^2}{2(1-e^{-\gamma t})}\right) .
\end{equation}
If, at the initial time, the quasi-particle was represented by the plane wave
$\psi(\bm{x}_0,0) \sim \exp\frac{i}{\hbar}\left(\bm{p}\cdot\bm{x}_0 \right)$, then at the time $t$ its state will be attenuated plane wave
\begin{equation}\label{E4.10}
    \psi(\bm{x},t) \sim e^{-\frac{3n\gamma t}{4}}\exp \left(i\left(\frac{\bm{p}}{\hbar}\cdot\bm{x}\right)  - i \omega t \right)
\end{equation}
with increasing frequency
\begin{equation}\label{E4.11}
    \omega = \frac{\bm{p}^2}{2\hbar m}\frac{e^{\gamma t}-1}{\gamma t}.
\end{equation}
It follows from (\ref{E4.8a})  and (\ref{E4.9}) that
\begin{equation}\label{E4.12}
    \mathcal{S}_F(\bm{x},t;\bm{x}_0,0) = \frac{\gamma m e^{-\gamma t}(\bm{x} - \bm{x}_0)^2}{2(1-e^{-\gamma t})}.
\end{equation}
Thus, in this case $\mathcal{S}_F$ coincides with classical action of the free \emph{S-particle}. Note, that matrix elements of $\hat{H}$ are decreased as
\[
    \langle \psi_{\bm{p'}}|\hat{H}|\psi_{\bm{p}} \rangle  \sim  \exp\left(-\frac{3n\gamma t}{2}\right) .
\]

\section{Discussion \label{Discus}}

In the previous sections, we have shown how to describe the evolution of systems when we have imprecise knowledge about the system's states and dynamics laws. The main assumptions, which are the basis for this approach, are: 1)-limit condition (\ref{E1.14}),(\ref{E1.14a}), 2)-causality principle in the fuzzy form and 3)-local topology of a system's state space.  We have shown that these assumptions inevitably lead to Hamiltonian (\ref{E2.10})-(\ref{E2.10b}) or Lagrangian equations of motion (\ref{E2.13})-(\ref{E2.13a}), if the system's evolution is described in $n+1$ dimensional space, where $n$ is the system's dimensionality, while the additional dimension can be considered as the mechanical action. Assumptions 2) and 3) are very general and can be applied to almost all systems, while in 1) it should be assumed that ``physical points" of the state space have a finite size and cannot be infinitely small. This means that the Lagrangian or Hamiltonian equations of motion could fail at a very small scale.

As mentioned above, the membership function $\mu(S,\bm{x},t)$ and possibility function $P(L,\bm{V}; S,\bm{x},t)$ are not equivalent to any kind of probabilities and, generally speaking, cannot be measured. In fact, concrete values of $\mu(S,\bm{x},t)$ and $P(L,\bm{V};S,\bm{x},t)$ for intermediate $0<\mu,P<1$  are ``human dependent" and only the values $\sup(\mu)=\sup(P)=1,\;\inf(\mu)=\inf(P)=0$ are objective, which reflects the nature of the fuzzy approach. As a rule, several experts will be consistent about the statements: ``\emph{At an initial time the system's state $\bm{x}_a$ is preferable to the state $\bm{x}_b$}'' or ``\emph{If the system is in state $\bm{x}_a$ then after a short time the state $\bm{x}_b$ will be preferable to the state $\bm{x}_c$}'', but if we will ask them to assign possibilities for each state or transition, they will come up with different numbers and the possibilities of one expert may not add up to those another. This means that  only correlation of preferability of events contains objective information in the description of a problem. The subjectiveness of the intermediate values of $\mu(S,\bm{x},t)$ and $P(L,\bm{V};S,\bm{x},t)$, however, does not present a serious problem. Indeed, the Master Equation (\ref{E1.17}) is covariant under the transformations
\begin{eqnarray*}
     \mu &\rightarrow& \Psi(\mu), \\
     P &\rightarrow& \Psi(P),
\end{eqnarray*}
if $T[P,\mu]$ is a \emph{consistent} t-norm \cite{MYbook}, meaning that
\[
   \Psi(T[P,\mu]) = T[\Psi(P),\Psi(\mu)],
\]
where$\Psi(x)$ is a monotonically increasing function with $\Psi(1)=1$ and $\Psi(0)=0$. It is obvious that under such transformations, only memberships grades of trajectories are changed, while the picture of the trajectories remains unchanged. Moreover, the most possible Lagrangian and picture of the most possible trajectories, $\mu(S(t),\bm{x}(t),t)=1$, remain unchanged for any choice of $T[P,\mu]$ and any ``expert's-dependent'' definition of intermediate values of $\mu$ .

The equations of motion (\ref{E2.10})-(\ref{E2.10b}) and (\ref{E2.13})-(\ref{E2.13a}) are more general than the usual Hamiltonian and Lagrangian equations in several ways. In some cases, they could be differential inclusions instead of differential equations, as they describe the dynamics of sets instead dynamics of points. Unfortunately, the mathematical apparatus of differential inclusions is poorly developed, and to date, we have only a few examples of explicit solutions of differential inclusions. Hence, this is much work to be done in this direction.

The second extension is \emph{dependence of the Lagrangian on action } ($S$-Lagrangian). In this case, the Lagrangian equations of motion acquire a non-zero right side, proportional to derivative of the $S$-Lagrangian with respect to the action. We have seen in section \ref{NonPhysSys} that even in simple cases, this leads to considerable changes in the system's behavior. Moreover, many features of the usual Lagrangian approach are lost. For example, the equations of motion with $S$-Lagrangian lose all additive integrals, they irreversible with respect to time and non-invariant under the addition to the Lagrangian of a function which is a total derivative with respect to time. On the other hand, this generalization allows one to define an $S$-Lagrangian for dissipative systems and systems which ``remember" their history.

It should be emphasized that our derivation of the equations of motion do not depend on any specific properties of the system or its Lagrangian. This means that equations (\ref{E2.10})-(\ref{E2.10b}), (\ref{E2.13})-(\ref{E2.13a}) and (\ref{E2.2c})-(\ref{E2.3}) give a reasonable method of applying the Hamiltonian or Lagrangian approach to non-physical systems. An example of such an application was presented in section \ref{NonPhysSys}.

Our brain is an unique object and can produce models of the world around us through ``perceptions" of the environment. Numerous sources in the literature and original experiments considered in \cite{MYbook} lead us to believe that the capability of perceiving takes place already on the neuron level and that the neural cell's processing of information could be close to fuzzy dynamics. The results of this paper give us some hint why the Hamiltonian or Lagrangian approach was so successful during the last centuries: the basic properties of this approach could be well compatible with the basic functioning of our``wet-ware" - neurons. Hence, ``background" processes of the brain (intuition) could work quite effectively when we use this way of thinking about the world.

The author acknowledge Dr.Tzvi Scarr for aid in editing the manuscript.

\appendix
\section{\label{A}}
In this appendix, we present an alternative form of the fuzzy dynamics equations. Let us consider the action as an additional dimension of the system's state space:
\[
    S=x_{n+1},\;\;\; L=V_{n+1},\;\;\; \frac{\partial}{\partial S}=\frac{\partial}{\partial x_{n+1}}.
\]
Equation (\ref{E2.1}) then takes the form of equation (\ref{E1.21}):
\begin{equation}\label{E.A1}
    \frac{\partial\mu}{\partial t} + (\bm{V}\cdot\nabla\mu)= 0,
\end{equation}
where $\bm{V}(\bm{x},\bm{k},t)$ should be found from
\[
  \lambda\frac{\partial P}{\partial \bm{V}} = \bm{k}.
\]
Here, $\bm{k}$ is a unit vector
\begin{equation}\label{E.A3}
    \bm{k} = \frac{\nabla \mu}{|\nabla \mu|},
\end{equation}
and the Lagrange multiplier $\lambda$ is found from
\[
  P(\bm{V};\bm{x},t) = \zeta_T(\mu).
\]

It has been shown in \cite{FS 1997}-\cite{MYbook} that the equations of motion corresponding to (\ref{E.A1}) are
\begin{eqnarray} \label{E.A2}
  \frac{d\bm{x}}{dt} &=& \bm{V}(\bm{x},\bm{k},t), \\ \label{E.A2a}
  \frac{d\bm{k}}{dt} &=& -\frac{\partial (\bm{k}\cdot\bm{V})}{\partial \bm{x}} +
  \bm{k}\left(\bm{k}\cdot\frac{\partial (\bm{k}\cdot\bm{V})}{\partial \bm{x}} \right).
\end{eqnarray}
In these equations, all dimensions are equivalent but (\ref{E.A2})-(\ref{E.A2a}) no longer have Hamiltonian form. It is easily seen, however, that equations (\ref{E.A2})-(\ref{E.A2a}) are equivalent to equations (\ref{E2.10})-(\ref{E2.10b}). Indeed, it is follows from (\ref{E2.9a}) and (\ref{E.A3}) that:
\begin{eqnarray*}
  k_{i\leq n} &=& -\frac{p_i}{\sqrt{\bar{1}+p^2}}, \\
  k_{n+1} &=& \frac{\bar{1}}{\sqrt{\bar{1}+p^2}}.
\end{eqnarray*}
where $\bar{1}$ is ``dimensional unit": $[\bar{1}]=[p]$. Therefore,
\begin{eqnarray} \label{E.A4}
  (\bm{k}\cdot\bm{V}) &=& \frac{\bar{1}}{\sqrt{\bar{1}+p^2}}\left( -(\bm{p}\cdot\bm{V}) +L \right) = - \frac{H}{\sqrt{\bar{1}+p^2}}
  \\ \label{E.A4a}
  \left(\bm{k}\cdot\frac{\partial (\bm{k}\cdot\bm{V})}{\partial \bm{x}} \right) &=& \frac{\bar{1}}{\sqrt{\bar{1}+p^2}}
  \left( \left(\bm{p}\cdot\frac{\partial H}{\partial \bm{x}}\right) +  \frac{\partial H}{\partial S}\right)= \frac{Q}{\sqrt{\bar{1}+p^2}}.
 \end{eqnarray}
Hence, (\ref{E.A2a}) can be written as
\begin{eqnarray*}
      -\frac{d}{dt}\frac{\bm{p}}{\sqrt{\bar{1}+p^2}} &=&\frac{\bar{1}}{\sqrt{\bar{1}+p^2}}\frac{\partial H}{\partial \bm{x}} -
      \frac{\bm{p}\,Q}{(\bar{1}+p^2)^{3/2}}  \;\;\;  (i\leq n) \\
      \frac{d}{dt}\frac{\bar{1}}{\sqrt{\bar{1}+p^2}} &=& \frac{1}{\sqrt{\bar{1}+p^2}}\frac{\partial H}{\partial S} + \frac{Q}{(\bar{1}+p^2)^{3/2}},
 \end{eqnarray*}
which immediately leads to Eq.(\ref{E2.10a}).

\section{\label{B}}

It follows from (\ref{E3.2}) that
\[
    \frac{\partial L_{mp}}{\partial v_i} = m(\rho - v)\frac{v_i}{v}.
\]
Assuming that $\rho$ is time-dependent ($\rho=\rho(\bm{x},t)$) and using (\ref{E2.26}), we can write
\begin{equation}\label{E.C1}
    \frac{d}{dt}\left(\frac{(\rho - v)v_i}{v}\right) - v\frac{\partial \rho}{\partial x_i} = - \gamma(\rho - v)\frac{v_i}{v}.
\end{equation}
Since
\[
    \frac{dv}{dt} = \frac{v_i}{v}\frac{dv^i}{dt} = n_i\frac{dv^i}{dt},
\]
and
\[
   \frac{d\rho(\bm{x})}{dt} = \frac{\partial\rho}{\partial t} + \frac{\partial\rho}{\partial x^i}v_i,
\]
we have
\begin{equation}\label{E.C2}
    \Phi_i^j\frac{dv_j}{dt} - \left(\delta_i^j - n_in^j\right)\frac{\partial \rho}{\partial x^j} =
    - (\partial_t\rho+\gamma(\rho - v))v_i,
\end{equation}
where
\[
   \Phi_i^j =(\rho - v)\left(\delta_i^j - \frac{\rho-v+v^2}{\rho - v}n_in^j\right).
\]
Multiplying both sides of (\ref{E.C2}) by
\[
    \Phi^{-1} = \frac{1}{\rho - v}\left(\delta_i^j - \frac{\rho-v+v^2}{v^2}n_in^j\right),
\]
we obtain
\[
   \frac{d\bm{v}}{dt} =  \frac{v^2}{\rho(\bm{x}) - v}\left(\nabla\rho - \bm{n}(\bm{n}\cdot\nabla\rho)\right) +
   \frac{\partial_t\rho + \gamma\rho(\bm{x}) - v)}{v}\bm{n},
\]
which leads to (\ref{E3.3a}) for $\partial_t\rho=0 $.

\section{\label{C}}

It is convenient to write $\bm{x}$ and $\bm{v}$ in the form  $\bm{x}=r\bm{e}$, $\bm{v}=v\bm{n}$, where $\bm{e}$ and $\bm{n}$ are unit vectors. In this case equations (\ref{E3.3})-(\ref{E3.3a}) take the form
\begin{eqnarray} \label{E.B1}
     \frac{dr}{dt}  &=& v(\bm{e}\cdot\bm{n}), \\  \label{E.B1a}
     \frac{dv}{dt}  &=& \gamma\frac{\rho(\bm{x}) - v}{v}, \\  \label{E.B1b}
     \frac{d\bm{n}}{dt}  &=& \frac{v}{\rho(\bm{x}) - v}\left(\nabla\rho - \bm{n}(\bm{n}\cdot\nabla\rho)\right), \\ \label{E.B1c}
     \frac{d\bm{e}}{dt}  &=& \frac{v}{r}\left(\bm{n} - \bm{e}(\bm{n}\cdot\bm{e})\right).
\end{eqnarray}
%00000000000000000000000000000000000
\begin{figure}
  \centering
  \includegraphics[width=12cm]{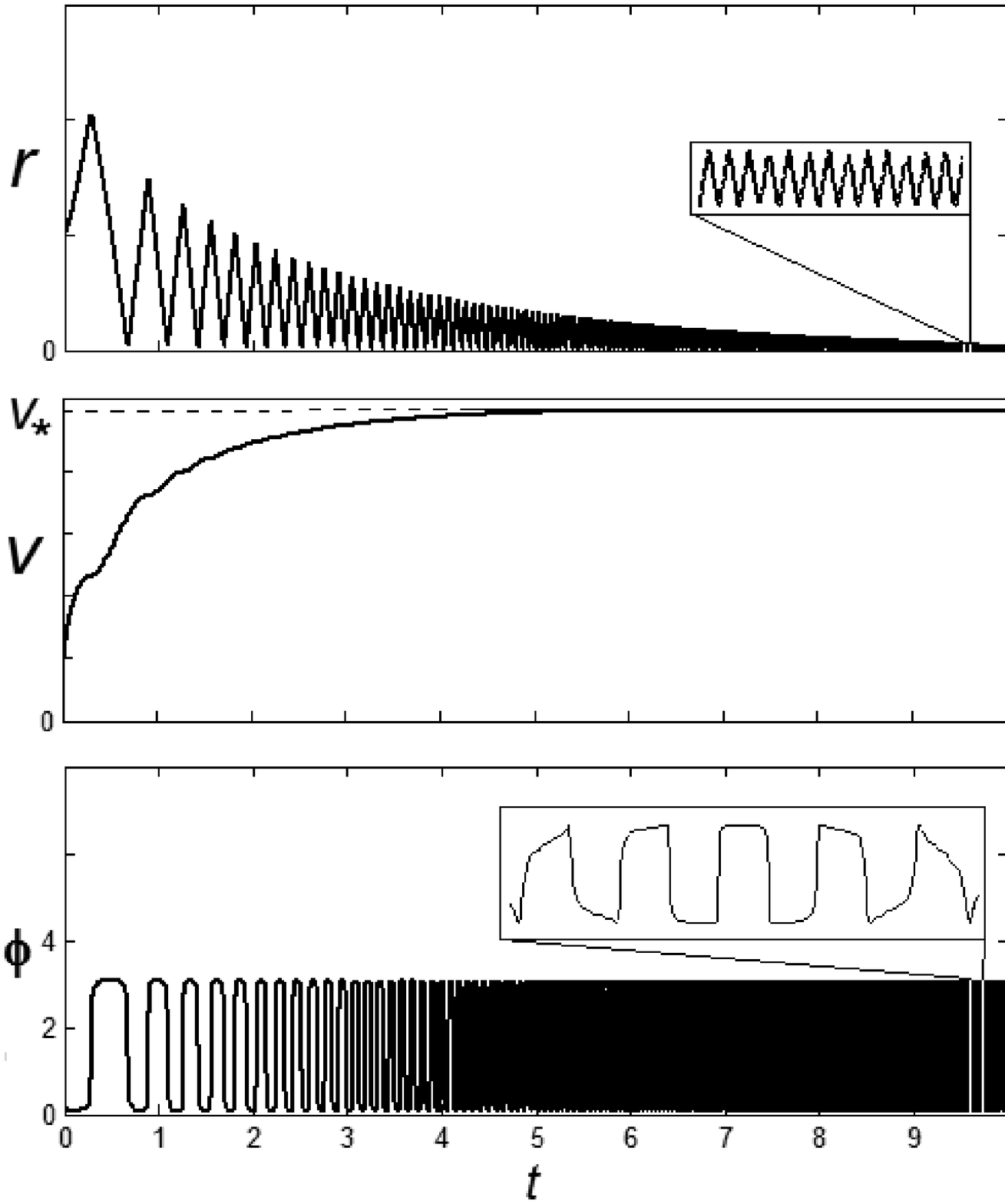} \\
  \caption{Behavior of $r(t)$,  $v(t)$ and $\phi(t)$ in equations (\ref{E.B2})-(\ref{E.B2b})  ($\gamma=5$).}
  \label{PSpect1}
\end{figure}
%00000000000000000000000000000000000
Assuming that the system is located in the vicinity of the maximum of $\rho(\bm{x})$: $\max\rho=\rho(0)$, where  $\rho(\bm{x})$  is almost axial symmetrical, we can put $\rho\simeq\rho(r)$ and $\nabla\rho = \rho\,' (r)\bm{e}$. On the surface, the vectors $\bm{e}$ and $\bm{n}$ can be written as
\[
    \bm{e} = \left[
                 \begin{array}{c}
                 \cos(\theta) \\
                 \sin(\theta) \\
                 \end{array}
                \right],\;\;\;\;\;
    \bm{n} = \left[
                 \begin{array}{c}
                 \cos(\varphi) \\
                 \sin(\varphi) \\
                 \end{array}
                 \right],
\]
so equations (\ref{E.B1})-(\ref{E.B1c}) can be rewritten in the form
\begin{eqnarray} \label{E.B2}
     \frac{dr}{dt}  &=& v\cos(\phi), \\  \label{E.B2a}
     \frac{dv}{dt}  &=& \gamma\frac{\rho(r) - v}{v}, \\  \label{E.B2b}
     \frac{d\phi}{dt}  &=& \frac{v}{r}\left(\frac{r\rho\,'(r) + \rho(r) - v}{\rho(r) - v}\right)\sin(\phi), \\ \label{E.B2c}
     \frac{d\theta}{dt}  &=& - \frac{v}{r}\sin(\phi),
\end{eqnarray}
where $\phi = \theta - \varphi$.

Equations (\ref{E.B2})-(\ref{E.B2c}) have a limiting circle satisfying
\begin{eqnarray} \label{E.B3}
     \rho(r_*) - v_* &=& r_*\rho\,'(r_*) + \rho(r_*) - v_* = 0, \\ \nonumber
    \phi_* &=& \frac{2n+1}{2} \pi, \;\;\; n=0,1, 2 ... \\ \nonumber
    \theta &=& - \frac{v_*}{r_*}t,
\end{eqnarray}
which leads to $r_* \to 0,\; v_*=\rho(0)$. Since the radius of the limiting circle tends to zero, the circle becomes unstable for any sampling of equation (\ref{E.B2b}). Indeed, near $r\ll 1$, we can write $v\simeq v_*=\rho(0)$, $\rho(r)\simeq \rho(0) - 0.5\rho\,''(0)r^2$, so (\ref{E.B2b}) can be written as
\[
   \frac{d\phi}{dt}  \simeq \frac{\rho(0)}{r}\left(\frac{r\rho\,'(r) + \rho(r) - \rho(0)}{\rho(r) - \rho(0)}\right)\sin(\phi),
\]
or
\[
    \frac{d\phi}{dt}  \simeq \frac{3\rho(0)}{r}\sin(\phi),
\]
which leads to the standard \emph{sinus-mapping}:
\begin{equation} \label{E.B4}
        \bar{\phi} \simeq \phi + K\sin(\phi),
\end{equation}
where
\[
    K \simeq \frac{3\rho(0)\delta t}{r}.
\]
Since, for any small $\delta t$, there is $r$ such that $K\gg 1$, the limiting circle is transformed to the strange attractor.

\section{\label{D}}

Consider an ordinary oscillator
\begin{eqnarray}\nonumber
         \frac{dx}{dt} &=& \frac{p}{m}, \\ \label{E.D1}
         \frac{dp}{dt} &=& -m\omega^2 x.
\end{eqnarray}
The function $P(\bm{V};\bm{x})$ in this case could be chosen as
\begin{equation}\label{E.D2}
    P(\bm{V};\bm{x}) = \Phi\left( \frac{(V_2 + m\omega^2 x)^2}{\sigma_1^2} +\frac{(V_1 - p/m)^2}{\sigma_2^2} \right),
\end{equation}
where $\Phi$ is the same as in (\ref{E2.14}). Now, let us consider $p$ as an $S$-variable. Then $\dot{p}=V_2=L$ will be our $S$-Lagrangian and
\begin{equation}\label{E.D3}
    P(L,v;S,x) = \Phi\left( \frac{(L + m\omega^2 x)^2}{\sigma_1^2} + \frac{(v - S/m)^2}{\sigma_2^2} \right).
\end{equation}
It follows from (\ref{E.D3}) and (\ref{E1.24}) that the $S$-Lagrangian is
\begin{equation}\label{E.D4}
    L = - m\omega^2 x +  \epsilon\sqrt{1 - \alpha^2(v - S/m)^2}.
\end{equation}
where $\epsilon=\sigma_1\sqrt{\Phi^{-1}(\zeta_T(\mu_0))}$ and $\alpha=\sigma_1(\epsilon\sigma_2)^{-1}$.
In accordance with (\ref{E2.9b}), we have
\begin{equation}\label{E.D5}
    \rho=\frac{\partial L}{\partial v} = -\frac{\epsilon\alpha^2(v-S/m)}{\sqrt{1 - \alpha^2(v - S/m)^2}},
\end{equation}
and the corresponding $S$-Hamiltonian is
\[
    H=v\frac{\partial L}{\partial v} - L = \frac{\rho S}{m} + m\omega^2 x-\frac{\sqrt{\epsilon^2 \alpha^2 + \rho^2}}{\alpha}.
\]
Thus, equations (\ref{E2.10})-(\ref{E2.10b}) take the form
\begin{eqnarray} \label{E.D6}
  \frac{dx}{dt} &=& \frac{\partial H}{\partial \rho} = \frac{S}{m} - \frac{\rho}{\alpha\sqrt{\epsilon^2 \alpha^2 + \rho^2}}, \\ \label{E.D6a}
  \frac{dS}{dt} &=& \rho\frac{\partial H}{\partial \rho} - H = - m\omega^2 x +
  \frac{\epsilon^2\alpha}{\sqrt{\epsilon^2 \alpha^2 + \rho^2}}, \\ \label{E.D6b}
  \frac{d\rho}{dt} &=& - \frac{\partial H}{\partial x} - \rho\frac{\partial H}{\partial S} =
  - m\omega^2 - \frac{\rho^2}{m},
\end{eqnarray}
Eq.(\ref{E.D6b}) has an obvious solution (for simplicity, we put $\rho(0)=0$ and $\varepsilon\alpha=m\omega$)
\begin{equation}\label{E.D6c}
    \rho = - m\omega \tan(\omega t),
\end{equation}
so Eqs.(\ref{E.D6})-(\ref{E.D6a}) take the form
\begin{eqnarray} \label{E.D7}
  \frac{dx}{dt} &=& \frac{S}{m} + \frac{\epsilon}{m\omega}\sin(\omega t), \\ \label{E.D7a}
  \frac{dS}{dt} &=&  - m\omega^2 x +\epsilon\cos(\omega t).
\end{eqnarray}
For $x(0)=x_0$ and $S(0)=0$  solution of (\ref{E.D7})-(\ref{E.D7a}) is
\begin{eqnarray} \label{E.D8}
       x     &=& x_0\cos(\omega t) + \frac{\epsilon t}{m\omega} \sin(\omega t), \\ \label{E.D8a}
       S     &=& - m\omega x_0\sin(\omega t)   + \frac{\epsilon}{\omega}[\sin(\omega t)  + \omega t\cos(\omega t)], \\ \label{E.D8b}
       H    &=& \frac{m\omega^2 x_0}{\cos(\omega t)} + \frac{\epsilon}{m}\left( \omega t\sin(\omega t) -
                    \frac{\sin^2(\omega t)}{\cos(\omega t)}\right).
\end{eqnarray}
We see that even for the most possible trajectory (for which $\epsilon = 0$) expressions (\ref{E.D8a}) and (\ref{E.D8b}) differ from the ordinary ones
\begin{eqnarray*} \nonumber
      H &=& \frac{m\omega^2 x_0^2}{2}, \\
      S &=& m\omega^2 x_0^2\sin^2(\omega t),
\end{eqnarray*}
while the most possible trajectory $x_{mp}(t)=x_0\cos(\omega t)$ is the same as for an ordinary oscillator: $x(t)=x_0\cos(\omega t)$.

\section{\label{Q}}

For $U(\bm{x})=0$ we will search solution of (\ref{E4.8})  in the form
\begin{equation}\label{Q.1}
    G_{\alpha}(\bm{x},t;\bm{x}_0,0) \sim e^{-n\gamma t/4}\exp\frac{i\alpha}{\hbar}\left( w(\alpha,t) + \frac{mu(\alpha,t)}{2}(\bm{x}-\bm{x}_0)^2\right)
\end{equation}
Substituting (\ref{Q.1}) to (\ref{E4.6}) one obtains
\begin{equation}\label{Q.2}
    \frac{\partial w}{\partial t} + \gamma \alpha\frac{\partial w}{\partial\alpha} + \gamma w + \left(\frac{\partial u}{\partial t} + \gamma \alpha\frac{\partial u}{\partial\alpha} + \gamma u  + u^2\right)\frac{m(\bm{x}-\bm{x}_0)^2}{2} = \frac{i\hbar n}{2\alpha}u,
\end{equation}
which leads to
\begin{eqnarray}\label{Q.3a}
  \frac{\partial w}{\partial t} + \gamma \alpha\frac{\partial w}{\partial\alpha} + \gamma w &=& \frac{i\hbar n}{2\alpha}u \\ \label{Q.3b}
  \frac{\partial u}{\partial t} + \gamma \alpha\frac{\partial u}{\partial\alpha} + \gamma u   &=& - u^2 .
\end{eqnarray}
These equations are easily solved and we obtain
\begin{eqnarray}\label{Q.4a}
  u(t) &=& \frac{\gamma u_0}{\gamma + u_0}e^{-\gamma t}\left(1-  \frac{u_0}{\gamma + u_0}e^{-\gamma t}\right)^{-1} \\ \label{Q.4b}
  w(t) &=& \mbox{const.} + \frac{i\hbar n}{2\alpha} \ln\frac{\gamma +  u_0(1-e^{-\gamma t}) }{\gamma u_0}.
\end{eqnarray}
In order to $\lim_{t\to 0} G_{\alpha}(\bm{x},t;\bm{x}_0,0) = \delta(\bm{x}-\bm{x}_0)$, we should specify
\begin{eqnarray*}
    u_0 &=& \infty \\
    \mbox{const.} &=& \frac{ n \hbar}{2i\alpha} \ln\frac{m\alpha}{2\pi i \hbar}
  \end{eqnarray*}
which immediately leads to (\ref{E4.9}).

%%%%%%%%%%%%%%%%%%%%%%%%%%%%%%%%%%%%%%%%%%%%%%%%%%%%%%%%%%%%%%%%%%%%%%%%%%%%%%%%%%%%%%%%%


\begin{thebibliography}{5}

\bibitem{Zadeh 1965} Zadeh LA, 1965, \textit{Inform. and Control} \textbf{8}, 338.

\bibitem{Zadeh 1975a} Zadeh LA, 1975,  Synthese \textbf{30}, 407.

\bibitem{Zadeh 1975b} Zadeh LA, 1975,  \textit{Inf. Scences} \textbf{8},199; \emph{ibid},  1975 ,301;
 \emph{ibid} ,  1975, \textbf{9}, 43.

\bibitem{Zadeh 2005} Zadeh LA, 2005,  \textit{Inf. Sciences}, \textbf{172}, 1.

\bibitem{Zadeh 2006} Zadeh LA, 2006,  \textit{Computational Statistics and Data Analysis}, \textbf{51}, 15.

\bibitem{Mizumoto 1989} Mizumoto M, 1989,  \textit{Fuzzy Sets and Systems}, \textbf{31}, 217.

\bibitem{Mesiar 2005} Klement EP, Mesiar R, 2005,  \textit{Logical, Algebraic, Analytic and Probabilistic Aspects of Triangular Norms}, (Elsevier).

\bibitem{Butnariu 1994} Butnariu D, 1994,  \textit{On Triangular Norm-Based Propositional Fuzzy Logic}, (\textit{Preprint} Haifa University, Haifa, Israel.)

\bibitem{Introduct 1} Hung T, Nguyen, Walker E, 2006,  \textit{A First Course in Fuzzy Logic} (CRC Press, London, NY).

\bibitem{Introduct 2} Bouchon-Meunier B, 1997,  \textit{Aggregation and Fusion of Imperfect Information}, (Springer-Verlag, NY)

\bibitem{Introduct 3} Novak V, Perfilieva I, Mockor J, 1999,  \textit{Mathematical Principles of Fuzzy Logic}, (Springer)

\bibitem{S 1994} Sandler U, 1994,   \textit{A New Approach to Fuzzy Logic}.  (In: Proc 12th IAPR Inter Conf on Pattern Recognition, Jerusalem, Israel)

\bibitem{FS 1997} Friedman Y, Sandler U, 1997,  \textit{Inter J of Chaos Theory and Application}, \textbf{2(3--4)}, 5.

\bibitem{MYbook} Sandler U. and Tsitolovsky L., 2008, \textit{Neural Cell Behavior and Fuzzy Logic}, (Springer, NY).

\bibitem{Aubin 1990}  Aubin JP, Cellina A, 1984,  \textit{Differential Inclusions}, (Springer-Verlag, NY)

\bibitem{Hullermeir 1997} Hullermeir E, 1997,  \textit{Int J Uncertainty Fuzziness and Knowledge-Based Systems}, \textbf{5}, 117.

\bibitem{Diamond 2000a} Diamond P, 2000,  IEEE Trans Fuzzy Systems \textbf{8}, 583.

\bibitem{F} Feynman R.P., 1965,  \textit{Quantum Mechanics and Path Integrals}, (McGraw Hill, NY).

\end{thebibliography}
\end{document}